\documentclass[onecolumn,showpacs]{revtex4}
\usepackage{amssymb}
\usepackage{makeidx}
\usepackage{amsmath}
\usepackage{graphicx}
\usepackage{epstopdf}
\usepackage{dcolumn}
\usepackage{bm}

\input{tcilatex}
\begin{document}

\title{X-ray absorption via electron-ion bremsstrahlung in Maxwellian plasma
at the exact consideration of Coulomb potential }
\author{A. G. Ghazaryan}
\email{amarkos@ysu.am}
\affiliation{Centre of Strong Fields, Yerevan State University, 1 A. Manukian, Yerevan
0025, Armenia}
\date{\today }

\begin{abstract}
Based on the results of exact consideration of electron-ion Coulomb
interaction, by numerical simulations for ultimate absorption coefficient,
we study the practically more interesting case of absorption of hard x-ray
quanta of frequency when photon energy is of the order of plasma
temperature, in high temperature plasma, within the high nuclear charges as
well (at which the Born approximation is not applicable), constituting the
temperature dependence of one-photon absorption coefficient. It is shown
that one can achieve the efficient absorption coefficient in these cases.
\end{abstract}

\maketitle



\section{Introduction}

With the advent of contemporary intense x-ray free electron lasers (FELs) 
\cite{1,2,3}, interest has grown to x-ray stimulated bremsstrahlung (SB)
absorption, because for the short laser pulses it may become the dominant
mechanism of the absorption of strong laser radiation in plasma \cite{4}.
Further, in the field of intense x-ray laser an electron in the SB process
may acquire the essential energy, which makes it effective mechanism for
laser-plasma heating, absorbing even one-two photons \cite{5}, \cite{6}. The
theoretical investigation of the SB process in gas or plasma medium in the
strong electromagnetic (EM) radiation field has been dedicated many papers
in various approximations over the scattering potential or EM wave field
using nonrelativistic \cite{4}, \cite{25,26,27,28,29,30,31,32,33a,34,35} or
relativistic descriptions \cite{36,37,38,39,40}. What concerns the
multiphoton absorption of plasma at high radiation intensities, the
relativistic theory of the latter has been succeeded in the Born \cite%
{41,42,43}, eikonal \cite{38}, and generalized eikonal \cite{44}
approximations over the scattering potential. Beyond this approximation for
the infrared and optical lasers in the multiphoton interaction regime, when
one could been applied classical theory and its main usable low frequency
approximation is considered in Refs. \cite{4}, \cite{30}, and it have been
generalized for the relativistic intensities of EM wave in Refs. \cite{39}, 
\cite{40}. It was shown in Ref. \cite{45}, where the interaction of strong
lasers with thin plasma targets of solid densities was investigated via
particle-in-cell simulations, that SB is dominant for electron densities
above 10$^{21}$\textrm{cm}$^{-3}$. The x-ray multiphoton absorption via
inverse bremsstrahlung absorption in such dense classical as well as quantum
plasmas in the present of strong wave field is investigated using quantum
consideration in Ref. \cite{5} in the Born approximation by the scattering
potential of a plasma ion, which demonstrates that for the relativistic
laser intensities the SB rate exhibits a tenuous dependence on plasma
temperature. So for middle wave intensities the consideration of dependence
on plasma temperature is of interest.

Meanwhile, for the optical or infrared lasers the quantum effects are small
because of the smallness of the photon energy. However for intense x-ray
radiation the nonlinear over the field quantum effects will be considerable.
What concerns the multiphoton absorption of plasma at high (and superhigh)
radiation intensities, the relativistic theory of the latter has been
succeeded major in the regime of small angles scattering for electrons SB 
\cite{5}, \cite{40}. In spite of this, and to find out the SB absorption
rates dependence versus the electrons temperature, in the present paper the
inverse-bremsstrahlung absorption of an x-ray laser field in the classical
Maxwellian plasma is considered by the quantum mechanical theory taking into
account a scattering potential field exactly, while the x-ray radiation is
treated by perturbation theory.

The earlier investigations of the one-photon absorption of high frequency
electromagnetic radiation by plasma via the stimulated bremsstrahlung of
electrons on the ion scattering potentials with the exact consideration of
Coulomb potential have been succeeded analytically only in two boundary
cases of photon frequency: for $\hbar \omega \ll \kappa T$ and $\hbar \omega
\gg \kappa T$ ($\kappa $ is the Boltzmann constant, $\omega $ is the EM wave
frequency) \cite{35}. Taking into account the significance of the problem
involving, specifically, the laser heating of plasma by current x-ray FELs
(Free Electron Laser) and its diagnostics, in the present paper, based on
mentioned results of exact consideration of electron-ion Coulomb
interaction, by further numerical simulations for ultimate absorption
coefficient, we study the practically more interesting case of absorption of
hard x-ray quanta of frequencies $\hbar \omega \sim kT$ in high temperature
plasma, within the high nuclear charges $Z$ as well (at which the Born
approximation is not applicable), constituting also the temperature
dependence of one-photon absorption coefficient.

The organization of this paper is as follows. In Sec. II the quantum
dynamics of x-ray absorption via electron-ion SB in Maxwellian plasma is
presented with analytical results for inverse-bremsstrahlung absorption
coefficient at the exact consideration of Coulomb field. In Sec. III, the
analytic formulas are considered numerically. Finally, conclusions are given
in Sec. IV.

\section{Basic theory of SB taking Coulomb potential exactly}

Let us investigate the inverse-absorption of x-ray radiation considering the
Coulomb potential exactly, while EM wave field is treated by perturbation
theory. The absorption coefficient $\alpha $ of the weak wave in the
one-photon approximation determined by the following formula:%
\begin{equation}
\alpha =\frac{\hbar \omega }{cE_{0}^{2}/8\pi }\dint \mathbf{f}(\mathbf{k}%
)[w_{a}(\mathbf{k})-w_{e}(\mathbf{k})]\frac{d^{3}\mathbf{k}}{(2\pi )^{3}}
\label{1b}
\end{equation}%
where $E_{0}$ is the amplitude of the electric field strength $\mathbf{E}(t)=%
\widehat{\mathbf{x}}E_{0}\sin \omega t$ of a linearly polarized EM wave in
dipole approximation with $\widehat{\mathbf{x}}\ $is the unit vector, $%
cE_{0}^{2}/8\pi $\ is the intensity of the EM wave in linear polarization
case, $w_{a}$ and $w_{e}$ total probabilities of one-photon absorption and
emission respectively, $\mathbf{f}(k)$ is the distribution function of the
electrons over the wavevectors $\mathbf{k=p/\hbar }$ , which is normalized
on the electron number density $N_{e}$\ as follows:%
\begin{equation*}
\dint \mathbf{f}(\mathbf{k})d^{3}\mathbf{k=}(2\pi )^{3}N_{e},
\end{equation*}%
$\varphi (r)=-Ze/r$ is the the Coulomb attractive potential, $e$\ is the
electron charge, $Ze$ is the nuclear charge. We using the results obtained
according to earlier analytic calculations \cite{35} for the isotropic
plasma with Maxwellian distribution function:%
\begin{equation}
\mathbf{f}(k)=N_{e}\left( \frac{2\pi \hbar ^{2}}{\mu \kappa T}\right)
^{3/2}\exp \left( -\frac{\hbar ^{2}\kappa ^{2}}{2\mu \kappa T}\right)
\label{2b}
\end{equation}%
($T$ is the temperature of electrons in plasma) in the case of hard
radiation 
\begin{equation}
a/k_{\omega }\ll 1.  \label{3b}
\end{equation}%
Here $a=Ze^{2}\mu /h^{2}$, $k_{\omega }=\left( 2\mu \omega /\hbar \right)
^{1/2}$. The absorption coefficient \cite{35} in lowest order of $%
a/k_{\omega }$ can been reduced to the form:%
\begin{equation}
\alpha =\alpha _{B}\sqrt{\pi }2^{3}\left( \frac{a}{k_{\omega }}\right) \chi
^{3/2}I\left( \chi ,\frac{a}{k_{\omega }}\right) ,  \label{3abc}
\end{equation}%
where $\chi =\hbar \omega /\kappa T$ is the electron-wave interaction
parameter, $N_{i}$\ is the ion number density, the both fields-dependent
function $I\left( \chi ,\frac{a}{k_{\omega }}\right) $ is%
\begin{equation*}
I\left( \chi ,\frac{a}{k_{\omega }}\right) =\dint\limits_{0}^{\infty }\exp
\left( -\chi t^{2}\right) \frac{\ln \left( t+\sqrt{1+t^{2}}\right) }{\exp
\left( \frac{2\pi a}{k_{\omega }t}\right) -1}
\end{equation*}%
\begin{equation}
\times \left[ 1-\exp \left( -\chi \right) \exp \left( \frac{2\pi a}{%
k_{\omega }t}\right) \right] dt,  \label{4b}
\end{equation}%
where%
\begin{equation}
\alpha _{B}=\frac{32\pi ^{2}N_{e}N_{i}}{3}\frac{Z^{2}e^{6}}{\mu c\hbar
^{2}\omega ^{3}k_{\omega }},  \label{5b}
\end{equation}%
is the absorption coefficient in the Born approximation, which is the same
in the cases for anisotropic electron distribution - monochromatic beam \cite%
{35}, as well for isotropic plasma with arbitrary momentum distribution of
electrons in the corresponding hard quantum case ($a/k_{\omega }\ll 1$) \cite%
{4}.

Previous analytical studies were successful in two boundary cases of photon
frequency. The first, for high-temperature plasma, $\kappa T\gg \hbar \omega
\gg \hbar ^{2}a^{2}/2\mu $, by expanding exponent $\exp \left( -\chi \right) 
$ in power series and performing integration in (\ref{4b}) one was obtained
the formula:%
\begin{equation}
\alpha =\alpha _{B}4\pi \left( \frac{a}{k_{\omega }}\right) \frac{\hbar
\omega }{\kappa T}.  \label{6b}
\end{equation}%
The second, for the case of $\hbar \omega \gg \kappa T$, when the absorption
coefficient in Maxwellian plasma was expressed by the following formula:%
\begin{equation*}
\alpha =\alpha _{B}4\pi ^{3/2}\left( \frac{a}{k_{\omega }}\right) \left( 
\frac{\hbar \omega }{\kappa T}\right) ^{1/2}\frac{id}{dx_{0}}\left\vert
F\left( 1-i\frac{a}{k_{\omega }},1,x_{0}\right) \right\vert ^{2}
\end{equation*}%
\begin{equation}
\times \left( 1-\exp (-2\pi a/k_{\omega })\right) ^{-1}\left( \exp (-2\pi
a/k_{T})-1\right) ^{-1},  \label{7b}
\end{equation}%
where $k_{T}=\sqrt{2\mu \kappa T}/\hbar $, $x_{0}=i4a/k_{\omega }$, $F\left(
z,1,x_{0}\right) $ is the confluent hypergeometric function, $%
x_{0}=i4a/k_{\omega }$. In the Born approximation by the scattering
potential, when $a/k_{\omega }\ll 1$ and $a/k_{T}\ll 1$, the expression (\ref%
{7b}) coincides with the expression (\ref{5b}).

Further we continue our study by numerical calculations for x-ray radiation
one-photon inverse-bremsstrahlung absorption coefficient.

\section{Numerical results for one-photon SB\ absorption coefficient of hard
x-ray}

To investigate the absorption coefficient $\alpha $\ by numerical
simulations we will utilize Eq. (\ref{3abc}). We study the dependence of
integral function $I\left( \chi ,\frac{a}{k_{\omega }}\right) $ (\ref{4b})\
versus parameter $\chi $. Besides, as follows from (\ref{3abc}), when
Coulomb potential is taken into account exactly the absorption coefficient
in the limit of low temperatures $a/k_{T}\gg 1$ decreases exponentially $%
\alpha \sim \exp \left( -2\pi a/k_{T}\right) $.(according to the formula (%
\ref{7b})). 
\begin{figure}[tbp]
\centering{\includegraphics[width=.51\textwidth]{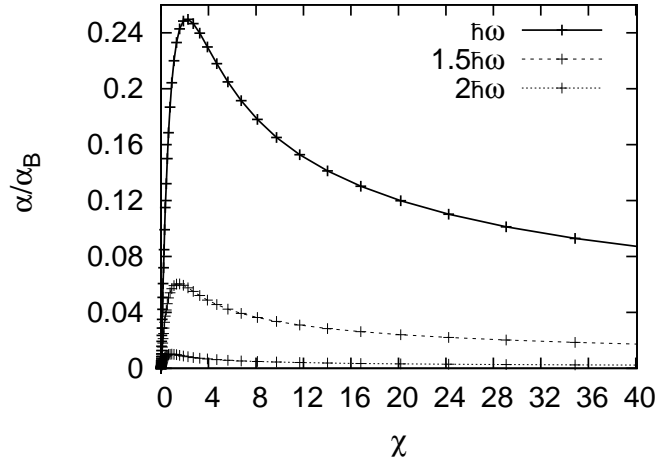}}
\caption{The absorption coefficient $\protect\alpha $ \ scaled to $\protect%
\alpha _{B}$ of one-photon hard radiation (of linear polarization) inverse
bremsstrahlung in Maxwellian plasma vs the dimensionless parameter $\protect%
\chi $ for $a/k_{\protect\omega }=10^{-5}$ and various photon energies.}
\label{11}
\end{figure}
\begin{figure}[tbp]
\centering{\includegraphics[width=.51\textwidth]{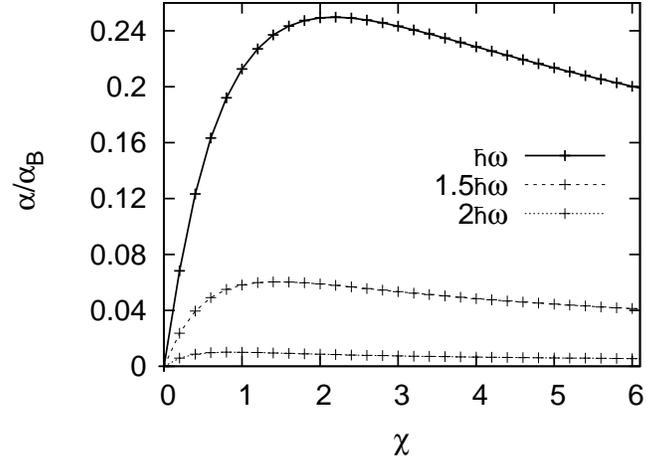}}
\caption{Same as Fig. 1 but for photon energies $\hbar \protect\omega \sim 
\protect\kappa T$.}
\label{22}
\end{figure}

\begin{figure}[tbp]
\centering{\includegraphics[width=.51\textwidth]{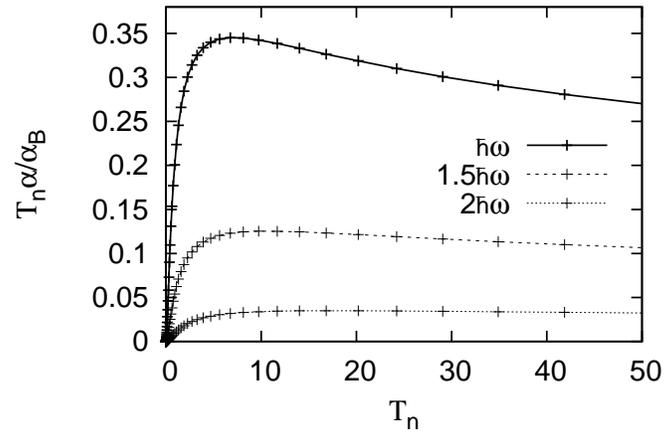}}
\caption{The absorption coefficient $\protect\alpha /\protect\alpha _{B}$ of
the one-photon hard radiation inverse bremsstrahlung absorption scaled to $%
1/T_{n}$\ vs the temperature $T_{n}$ for setup of Fig. 1.}
\label{33}
\end{figure}

The magnitude of the electron momentum change during the scattering process
of low-temperature electrons in the Coulomb potential is much less than is
necessary for real absorption of quantum with energy many times greater than
the electron's energy $\kappa T$.

\begin{figure}[tbp]
\centering{\includegraphics[width=.51\textwidth]{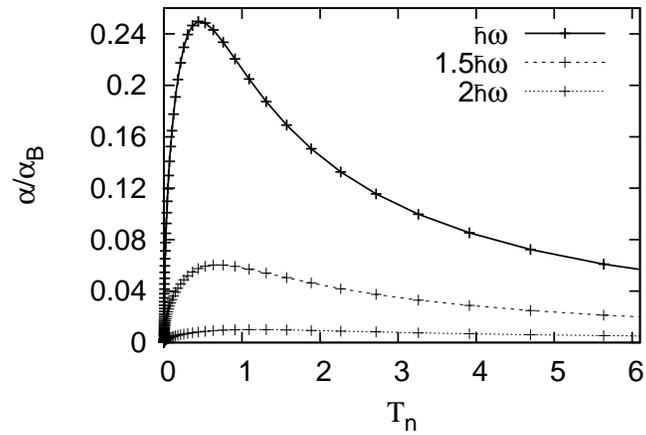}}
\caption{Same as Fig. 3 but for the temperature $T_{n}\sim 1$.}
\label{44}
\end{figure}

We study the practically more interesting case of absorption of hard x-ray
quanta, when $a/k_{\omega }\ll 1$, making integration by codes using Gauss'
rule. To show the dependence of the one-photon inverse-bremsstrahlung rates
on the photon energy, in the Fig. 1 the absorption coefficient $\alpha $
scaled to $\alpha _{B}$ (\ref{5b}) of linearly polarized wave in Maxwellian
plasma versus ${\chi }$ for various photon energies is shown ($a/k_{\omega
}=10^{-5}$). As seen from this figure, the SB coefficient is considerably
suppressed with the increase of the wave quanta energy,\ and in the case of
hard radiation in high temperature plasma with the high nuclear charges $Z$
(beyond the Born approximation) the inverse-bremsstrahlung absorption
coefficient dependence on ${\chi }$ (and plasma temperature) has the
characteristic maximum after which it decreases with increasing of the
parameter $\chi $. In Fig. 2 the absorption coefficient $\alpha $ scaled to $%
\alpha _{B}$ dependence versus parameter $\chi $ is showed for the
frequencies $\omega \sim \kappa T/\hbar $ ($T_{n}\equiv T/\hbar \omega \sim
1 $), which demonstrated that as the energy of the quantum of the wave
increases, the probability of SB\ absorption reaches the maximal value when $%
\hbar \omega =\kappa T$.

To compare with the case when the electron-ion interaction is considered in
the Born approximation \cite{4}, in Fig. 3 the dependence of the one-photon
hard radiation inverse-bremsstrahlung absorption coefficient $\alpha /\alpha
_{B}$ scaled to $1/T_{n}$ versus the plasma temperature is shown for various
laser radiation frequencies and for small $a/k_{\omega }$ (for linearly
polarized wave), and Fig. 4 shows the same in the vicinity of $\hbar \omega
=\kappa T$. Fig. 3 demonstrates that when the electron-ion interaction is
taken into account exactly the absorption coefficient of high temperature
plasma at $\kappa T\gg \hbar \omega \gg \hbar ^{2}a^{2}/2\mu $ (${\chi \ll 1}
$) dependence versus the plasma temperature $T_{n}$ is suppressed. Thus, now
obtained results revealed a different dependence of absorption coefficient
on the plasma temperature $T$, in contrast to the known results obtained in
the Born approximation (\ref{5b}). If the Coulomb potential is taken into
account exactly the absorption coefficient of hard radiation depends on
plasma electron temperature as $\alpha \sim 1/T$, it decreases more slowly
with plasma temperature, than in the Born approximation by the electron-ion
interaction ($\alpha \sim \ln T/T^{3/2}$) \cite{4}. As it is showed in Fig.
4, for hard radiation in the case of frequencies $\hbar \omega \sim \kappa T$
in high temperature plasma with the high nuclear charges $Z$ one-photon
absorption coefficient dependence versus $T$\ has the maximum. The numerical
simulations give the characteristic curve for the dependence of $\alpha
/\alpha _{b}$\ versus plasma temperature.

\section{Conclusion}

We have presented the numerical investigation of x-ray radiation one-photon
SB\ absorption coefficient under the limits on the main characteristic
coefficients of the process, such as parameter ${\chi }$ and plasma
temperature, based on earlier results of the exact quantum mechanical
consideration of the dynamic of electron-ion Coulomb interaction in the
presence of a transverse EM wave, when Coulomb potential is taken exactly,
while the EM wave is treated by perturbation theory. The coefficient of SB
absorption has been calculated considering the classical Maxwellian
distribution. We study the more interesting case of absorption of hard
x-rays ($a/k_{T}\ll 1$, ${\chi \sim 1}$) in high temperature plasma with
high nuclear charges as well, at which the Born approximation by scattering
potential is broken. The obtained results demonstrate that one can achieve
the efficient absorption coefficient in this case, and the SB rate is
suppressed with increase of the parameter ${\chi }$ and plasma temperature,
and wave quanta energy. For large values of $T$, the absorption coefficient $%
\alpha $ decreases as $1/T$ in contrast to the case, when the scattering
potential considering as perturbation where one has the dependence $\ln
T/T^{3/2}$ \cite{4}. In the limit of low plasma electron temperatures, $%
a/k_{T}\gg 1$, one-photon SB\ absorption coefficient exponentially decreases
because of the classical nature of the electron interaction with the Coulomb
potential, whereas it interaction with EM wave has quantum nature
(one-photon inverse bremsstrahlung). The problem is significant connected
with the laser heating of plasma by current x-ray FELs and its diagnostics.

\begin{acknowledgments}
The author is deeply grateful to the Prof. Hamlet K. Avetissian for
permanent discussions during the work on the present paper, for valuable
comments and recommendations. This work was supported by the RA MES State
Committee of Science.
\end{acknowledgments}

\end{document}